 \documentclass [aps,pra, onecolumn,a4paper]{revtex4}

 \pdfoutput=1

\usepackage{amssymb,amsmath,amsfonts,graphicx}
\usepackage{bm}
\usepackage{epsfig,color,times}
 \usepackage{bm,color}
\usepackage{graphicx}
\usepackage{amssymb}
\usepackage{epstopdf}

\begin{document}

\title{Turbulent plane Poiseuille flow}
\author{Yves Pomeau and Martine Le Berre}
\affiliation{ 
Laboratoire d'Hydrodynamique, Ladhyx, (CNRS UMR 7646), Ecole Polytechnique, 91128 Palaiseau, France }

\date{\today}

\begin{abstract}
The ultimate goal of a sound theory of turbulence in fluids is to close in a rational way the Reynolds equations, namely to express the time averaged turbulent stress tensor as a function of the time averaged velocity field.  This closure problem is a deep and unsolved problem of statistical physics whose solution requires to go beyond the assumption of a homogeneous and isotropic state, as fluctuations in turbulent flows are strongly related to the geometry of this flow. This links the dissipation to the space dependence of the average velocity field. 
Based on the idea  that dissipation in  fully developed turbulence is by singular events resulting from an evolution described by the Euler equations, it has been recently observed that the closure problem is strongly restricted, and that it implies that the turbulent stress is a non local function (in space) of the average velocity field, an extension of classical Boussinesq theory of turbulent viscosity.  The resulting equations for the turbulent stress are derived  here in one of the simplest possible physical situation, the turbulent Poiseuille flow between two parallel plates. In this case the integral kernel giving the turbulent stress, as function of the averaged velocity field, takes a simple form leading to a full analysis of the averaged turbulent flow in the limit of a very large Reynolds number. In this limit one has to match a viscous boundary layer, near the walls bounding the flow, and an outer solution in the bulk of the flow. This asymptotic analysis is non trivial because one has to match solution with logarithms. A non trivial and somewhat unexpected feature of this solution is that, besides the boundary layers close to the walls, there is another "inner" boundary layer near the center plane of the flow.  

\end{abstract} 

\maketitle

\section{Introduction}
\label{Introduction}
In a recent work we introduced a new way of modeling turbulent flows in "real situations", that is where turbulence is due to an incompressible flow at large Reynolds number in given geometries.
 This was partly for the purpose to show how to use "concretely" the idea that, in such a turbulent flow, the turbulent stress tensor (RST) can be described by explicit expressions of  the global (in the sense of space dependent and time-averaged) velocity field. These expressions are
  strongly constrained by the fact that the dissipation in such a turbulent flow does depend only on the parameters of the local average velocity field. As shown in ref \cite{1} this leads quite naturally to write the turbulence stress as a non local quantity depending quadratically on the velocity field and its space derivatives, with non diagonal components of the RST of the form
 \begin{equation} 
 \tilde \sigma_{ij}^{Re}({\bold{x}}) = \gamma \rho \vert  {\bold{\nabla}}\times {\bold{u}}({\bold{x}})\vert^{1-\alpha} \int {\mathrm{d}} {\bold{x}}' \, \vert {\bold{\nabla}}   \times {\bold{u}}({\bold{x'}})\vert^{\alpha}  \, {\bold{K}}({\bold{x}},{\bold{x'}})\,
  (u_{i, j} + u_{j,i})({\bold{x}}') 
 \label{eq:sig11}
\end{equation} 
where  ${\bold{u}}({\bold{x}})$ is the  time average of the velocity, the  exponent $\alpha$ is such that $0 <  \alpha <1$, $\tilde \gamma$  is a dimensionless constant, and $K({\bold{x}},{\bold{x'}})$ is the Green\textquoteright s function of the Laplace operator with Dirichlet boundary conditions.
This allowed a detailed analysis of the turbulent mixing layer, that led to an almost fully explicit solution in the case of a relatively small velocity difference between the two parallel flows merging in the wake of the splitting plate. However  in the mixing layer set up, the solid boundaries play only a minor role, contrary to most real turbulent flows. Therefore we thought it of interest to use the same theory in a turbulent flow where solid boundaries play instead a fundamental role.  Here we study a classical example of such a flow,  namely the Poiseuille flow between two parallel planes. In such a flow one has first to write the integral kernel that enters into the expression of the RST. Generally this integral kernel is the Green's function of the Laplacian with Dirichlet boundary condition. In this geometry it is quite elementary to derive this integral kernel. The Dirichlet boundary condition plays a central role in this analysis. Thanks to it, the turbulent stress decays smoothly to zero as one approaches the boundary, that constitutes one of the fundamental assumptions in Prandtl theory leading to a  log-dependence law for the profile. Compared to this well known theory however the present work deals with equations not limited to the neighborhood of the wall. Therefore it is possible to do the full matching between the viscous sublayer and the flow far from it. This is no trivial matter because of the occurrence of  another logarithm in the boundary layer solution. 

Our detailed numerical analysis has shown also a rather unexpected phenomenon, namely the existence of another boundary layer near the center line, far from the wall. This is a boundary layer in the sense that, without viscosity, the slope of the velocity is discontinuous in the center of the channel, see Sec.\ref{Matching 2}.
 Once the viscosity is taken into account, this discontinuity becomes a narrow interval of width of the order of the inverse Reynolds number, where the slope of the velocity, as a function of the transverse position, goes continuously from a positive to a negative constant. Physically this corresponds to a narrow jet near the center line. 

In section \ref{statement} we define  the geometry of the problem and give the equations to be solved together with their boundary conditions. 
In section \ref{Simplifications} we write  first the equation for the balance of momentum in the simplest possible form and solve it near the wall to get the log-law of the wall, then 
 we carry explicitly the matching of this boundary layer solution with the solution in the bulk, namely outside of the close vicinity of the wall, and finally we study
 the  viscous smoothening of the  discontinuous slope observed numerically in the center of the fluid layer. 
Section \ref{turbulentdrag} is more speculative than the previous ones. We try to draw some conclusions from the analysis of the previous sections concerning the general question of the turbulent drag on blunt bodies, a question with an history going back to the Principia. Our point is that there are two kinds of drag forces, the "skin drag" with a  friction coefficient decaying like inverse of the log of the Reynolds number  
and another drag,  named turbulent drag here, formed when the flow  has to round  an obstacle. The turbulent drag is the one found by Newton,  it is   proportional to the square of the velocity with a coefficient independent on the Reynolds number.

\section{Statement of the problem}
\label{statement}

    \begin{figure}
\centerline{ 
\includegraphics[height=2in]{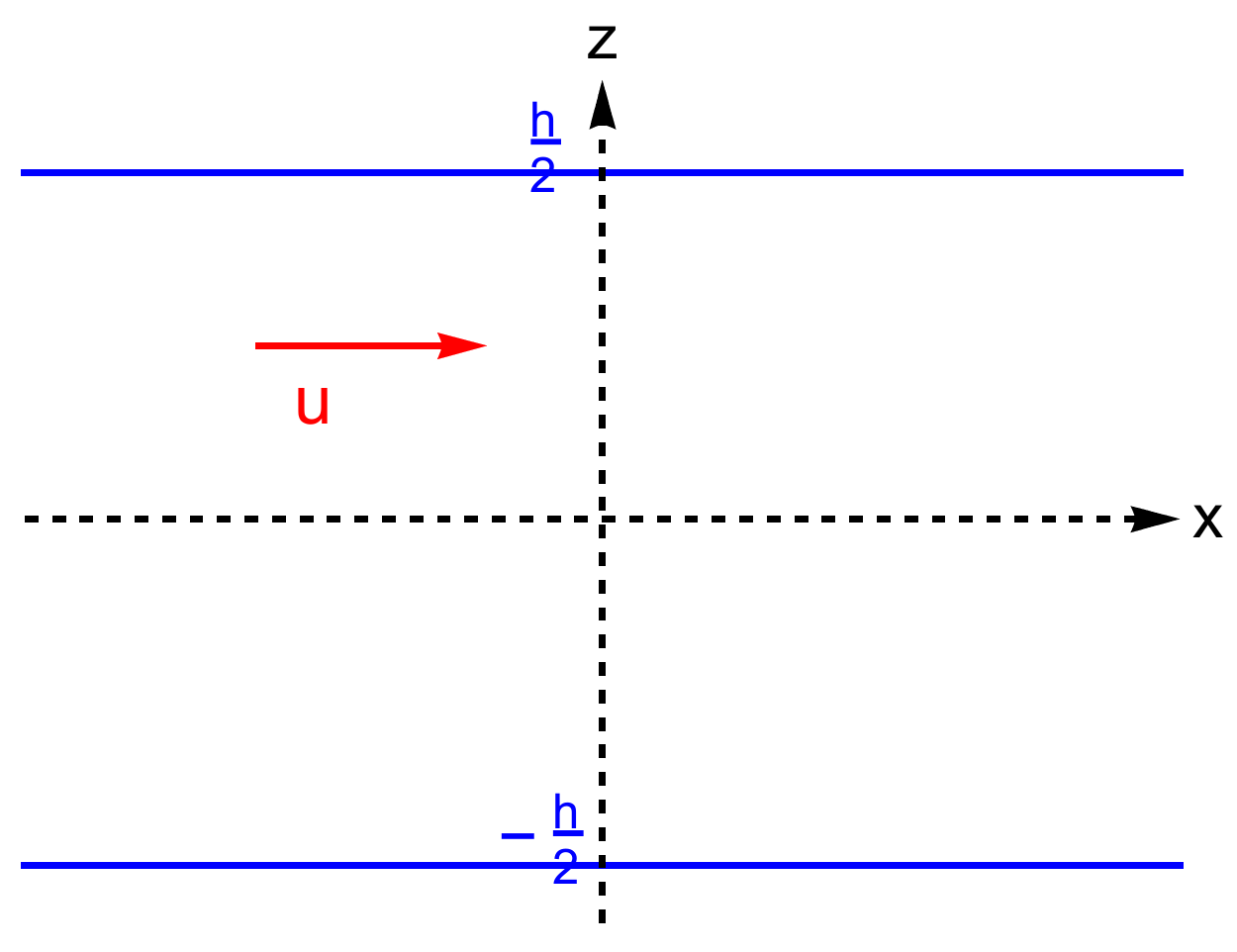}
  }
\caption{ Poiseuille plan set up.}
\label{fig:scheme}
\end{figure}

The geometry under consideration is a turbulent plane Poiseuille flow in between two parallel plates  located respectively at the elevation $ z = -h/2$ and $z = h/2$, see Fig.\ref{fig:scheme}.  Because of the symmetry with respect to the middle-plane located at $z=0$, the time average velocity ${\bold{u}}$  has a single component, $u$, along the $x$  axis, which depends on $z$ which is therefore the  relevant variable.  The other dimension associated to the $y$ coordinate will play no role at all.  The flow is driven by a constant uniform pressure gradient $g$ along $x$. Lastly the kinematic viscosity of the fluid is $\nu$ and its mass density is $\rho$. Our equation for the time averaged Reynolds turbulent stress (the only one we shall deal with) includes the pressure gradient $g=\frac{\mathrm{d} p(z)}{\mathrm{d} z}$ and the component $\sigma_{xz}$ of this stress tensor
\begin{equation} 
\sigma_{xz}=\rho <u'_{x} u'_{z}>
 \label{eq:sig0}
\end{equation}  
where  $u'_{i}$  is the the fluctuating part of the velocity component along the i-coordinate. On average the stress balance along $x$ yields the equation
\begin{equation} 
\frac{1}{\rho}\frac{\mathrm{d} \sigma_{xz}} {\mathrm{d} z} - \nu \frac{\mathrm{d}^2 u(z)}{\mathrm{d} z^2} =  -\frac{g}{\rho }
 \label{eq:sig1}
\end{equation}  
or, after  integration over the variable $z$, 
\begin{equation} 
 \frac{\sigma_{xz}}{\rho}  - \nu \frac{\mathrm{d}u(z)}{\mathrm{d} z} = \frac{\vert g \vert}{\rho} z
 \label{eq:sig1b}
\end{equation}  
where  $\vert g \vert = - g $ because  we assume that the pressure gradient is negative.

Note that we use straight symbol for the derivatives because all functions involved depend on the variable $z$ only. Without the contribution of the turbulent stress, this equation reduces  to the one of the laminar plane Poiseuille flow when the boundary conditions $u(z) = 0$ are imposed for $ z =  \pm h/2$.  What we shall do now is to look at the solution of this equation in the opposite limit where  $ \nu$ is small, which is of course equivalent  to the limit of a very large Reynolds number. This requires to express the turbulent stress $ \sigma_{xz}$ as a function of the mean velocity field $u(z)$ then to solve the  equation with the boundary condition, an implicit problem.

As stated in the introduction, our model  (\ref{eq:sig11}) for the RST satisfies 
various symmetries based on the fact that dissipation is due to singular events described by solutions of the Euler fluid equations.  
In the geometry under consideration, 
the equation  (\ref{eq:sig11}) for  $\sigma_{xz}$  becomes
 \begin{equation} 
 \sigma_{xz}(z)/\rho = \gamma \vert  \frac{\mathrm{d} u(z)}{\mathrm{d} z} \vert^{1-\alpha} \int_{-h/2}^{h/2} {\mathrm{d}} z' \vert  \frac{\mathrm{d} u(z')}{\mathrm{d} z'} \vert^{\alpha} K(z, z')   \frac{\mathrm{d} u(z')}{\mathrm{d} z'}
 \label{eq:sig2}
\end{equation}  
without integration constant because we assume that the velocity field $u(z)$ is an even function. In the expression (\ref{eq:sig2}) $\gamma$ is a numerical constant of order $1$, 
and  various simplifications appear compared to the expressions given in  \cite{1} and \cite{2}. First the vorticity has been written here as the derivative $ \frac{\mathrm{d} u(z)}{\mathrm{d} z}$, as it follows from the geometry. Furthermore the integral is only carried over the variable $z$. As explained in \cite{2} this leads to take for $K(z, z') $ the Green's function of the Laplacian in the space of the relevant variable with the Dirichlet boundary condition on the solid surfaces. In the present case this leads to the following equation for $K(z, z')$ 
 \begin{equation} 
 \frac{\mathrm{d^2} K(z, z')}{\mathrm{d} z^2} =  \delta(z - z')
  \label{eq:sig3}
\end{equation}  
where $\delta(.)$ is Dirac's delta function. 

Because the equation  (\ref{eq:sig1}) is posed with $z$ as variable, the solution satisfies the Dirichlet boundary condition $K(z, z') = 0$ for $z =  \pm h/2$ and so yields a stress tensor  $\sigma_{xz}(z)$ vanishing for $z =  \pm h/2$, an important constraint on Reynolds turbulent stress, expressing that the component of the velocity normal to the solid boundary is zero, both for the average velocity and its fluctuations $u'_i$ ($i$ component index) that enters into the Reynolds stress as $  \sigma_{xz}(z) = <u'_x u'_z>$. Here $u'_z = 0$ for $z =  \pm h/2$. 
The solution of Equation  (\ref{eq:sig3}) with the Dirichlet boundary condition at $ z =  \pm h/2$ is
 \begin{equation} 
K(z, z') =  \frac{1}{2} (\vert z - z' \vert + \frac{2}{h} zz'- \frac{h}{2})
  \label{eq:sig4}
\end{equation}  
The function $K(z, z') $ has two properties that will play a role later, first it is an even function under the joint change of sign of $z$ and $z'$, 
 \begin{equation}
  K(z, z') = K(-z, -z') ,
  \label{eq:Ksym}
\end{equation}  
Furthermore $K(z, z')$  is a continuous function of $z$ and $z'$ and its  derivative with respect to $z$ has a finite jump of $1$ at $ z = z'$. The integral over $z'$ in (\ref{eq:sig2}) can be written as
 \begin{equation} 
L(z)=\int_{-h/2}^{h/2} {\mathrm{d}} z' K(z, z') f(z') 
  \label{eq:L}
\end{equation} 
where
 \begin{equation}
f( z') =  \vert  \frac{\mathrm{d} u(z')}{\mathrm{d} z'} \vert^{\alpha}  \frac{\mathrm{d} u(z')}{\mathrm{d} z'} ,
  \label{eq:f}
\end{equation} 
and can be simplified a bit;  first we note that the last term in $K$, namely $ -  \frac{h}{2}  $, doesn't contribute  to the odd integrant. The remaining part of the integral gives
 \begin{equation} 
L(z)= \frac{1}{2}\int_{-h/2}^{h/2} {\mathrm{d}} z' f(z')    ( \vert z - z' \vert + 2 \frac{z z'}{h})
  \label{eq:sig5}
\end{equation}  
 From equation (\ref{eq:sig1}), $u(z)$ is an even function of $z$, and $  \sigma_{xz}(z) $ an odd function of $z$. In the limit of large Reynolds number, the solution of this equation, which is odd with respect to $z$, is
 \begin{equation} 
 \sigma_{xz}(z) = \vert g \vert z
   \label{eq:nu0}
\end{equation}  

Obviously $ \sigma_{xz}(z)$, as given by this expression, does not vanish for $ z =   \pm h/2$, so that some (non trivial) changes must be made to get a solution satisfying the boundary condition  $\sigma_{xz}(\pm h/2) = 0$.

The scaling laws to be derived from equation  (\ref{eq:sig1}) have to relate the order of magnitude of the quantities with a physical meaning, here the average velocity $u$, to the parameters of the problem , namely $g$, $h$ and $\nu $. Taking $h/2$ as length scale one gets rid  of any quantity with a physical dimension in the problem by taking 
 \begin{equation} 
u_{*}= (\frac{\vert g \vert h}{2\rho})^{1/2}
   \label{eq:ustar}
\end{equation}  
as unit velocity, (an idea going back to the second half of the eighteenth century and due to the French engineer Ch\'ezy  \cite{3} with $g/2\rho$ replaced by the slope of the bottom of a river times the acceleration of gravity and $h$ a length depending on the depth and width of the flowing river). Taking this scaling law one makes appear instead of $\nu$, the inverse  of a Reynolds number 
 \begin{equation} 
 \frac{1}{Re}\approx  \frac{2\nu}{ u_{*} h}. 
   \label{eq:Re}
\end{equation}  
  although $Re$ will be defined later  by the relation (\ref{eq:Re2}). We are looking for the solution of equation (\ref{eq:sig1}) in the limit of a very large Reynolds number, namely for a small viscosity. Therefore it is natural in this limit to assume first that the term of viscous stress in equation (\ref{eq:sig1}) is negligible. Neglecting this term leads to a parameterless equation where all terms have formally the same order of magnitude. The solution $u(z)$ we are looking for is an even function of $z$ as we shall see. However it has no reason to satisfy the boundary condition $u = 0$ for $z =  \pm h/2$ (this condition is imposed at $z =  \pm 1/2$ is one takes $h$ as unit length).  This situation is actually fairly common in fluid mechanics where the boundary condition for the velocity is different in perfect fluids and in viscous fluids. At high Reynolds number this makes the viscosity significant only in a thin layer near the solid surface where the tangential velocity drops from a finite value far from the layer to zero on the solid surface.  There is an added complexity in this problem, because there is another boundary layer at the center of the cell, at $ z = 0$, as discussed in section \ref{Matching 2}. 
  
We can derive a simpler expression of the integral in equation  (\ref{eq:sig5}), by using various symmetries of the integrant. Let us introduce the function $F(z)$ such that 
  \begin{equation} 
 F(z) = \int_{0}^{z} {\mathrm{d}} z' f(z')  
  \label{eq:Gz}
\end{equation}  
The function $F(z)$ is an even function of $z$. Integrations by part yield 
  \begin{equation} 
L(z) =  \int_{-h/2}^{h/2} {\mathrm{d}}z'  K(z,z') f(z')   =   \frac{1}{2}  \int_{-z}^{z} {\mathrm{d}} z' F(z') -\frac{z}{h} d
  \label{eq:sig5.2}
\end{equation}  
where 
 \begin{equation} 
d  =   \int_{-h/2}^{h/2} {\mathrm{d}} z' F(z') 
  \label{eq:d}
\end{equation}

\section{Solution}
\label{Simplifications}

This question of the boundary layer has been discussed for a long time and remains tricky because of the occurrence of logarithms in the solution. The present approach is in principle more straightforward than some others, because it relies on a fully explicit equation for the stress valid all the way in the  turbulent flow, from the  wall until the bulk.

The difficulty in this problem is twofold. First one has to solve the equation for $u(z)$ in the turbulent domain without the viscosity term and then to match this solution with a boundary layer where viscosity plays a role. The first problem, (average velocity in the turbulent domain far from the wall in a sense to be made precise) is already non trivial because it relies on the solution of a parameterless  non linear  integro-differential equation. Even though one can only hope to get a numerical solution, one needs at least to have some information on this solution, particularly near $\vert z \vert  = h/2$ where it has to be matched with the one in the boundary layer, this one depending on the viscosity. 

Below we solve first the problem close to the wall, then away from it, and match finally the two solutions. We show that the full solution agrees with the observation that the friction coefficient (namely the dimensionless coefficient in the Ch\'ezy formula for the average speed as a function of $g$) tends to zero logarithmically as the Reynolds number increases. This (non trivial) property is special to pipe flows with infinitely smooth boundaries. As discussed in Sec. \ref{turbulentdrag} below, there is no such a  logarithmic decrease of  the friction (with respect to the Reynolds number) in the $C_x$ coefficient of Newton's quadratic law for the drag of blunt bodies at large speed; we explain why the $C_x$ coefficient tends to a constant at infinite Reynolds number. 

\subsection{Boundary layer solution}
\label{sec:wall}

We shall prove that the (local) linearity of $L(.)$ with respect to the distance to the wall will ultimately yield the "log-law of the wall", but contrary to the standard derivation of this law, we do not have to \emph{assume} that the turbulent stress is proportional to the distance to the wall, but we \emph{derive} this from an explicit expression for the stress $ \sigma_{xz}(z)$. The latter expression  is a priori valid near the wall and far from it, and satisfies the boundary condition that $ \sigma_{xz}(z) = 0$ on the wall just because the velocity field, average and fluctuation, has no component normal to the wall. We emphasize that the boundary layer exists because the solution of equation (\ref{eq:sig1b}) has to satisfy the condition of vanishing contribution of Reynolds stress at the boundaries.

Near $ \vert z \vert  =  h/2$,  the first term in the Taylor expansion  of $L(z)$   is  
 \begin{equation} 
 L(z) _{z  \to  \pm \frac{h}{2}}  \approx \pm ( \frac{h}{2}-\vert z \vert) \left(\frac{d}{h} - F(\frac{h}{2})\right)
  \label{eq:sig5.3}
\end{equation}
which is linear with respect to the distance to the wall, and vanishes at $\vert z \vert=h/2$. We notice that
it depends on the whole solution via $d$, defined in (\ref{eq:d}). This is a physical consequence of the underlying theory which does not assume the existence of any scaled length besides the width of the channel outside the close neighborhood of the wall.  It follows that  the Reynolds stress tensor depends on turbulent fluctuations existing in the whole fluid domain and cannot be seen as depending on local quantities only. Said otherwise, the  implicit expression (\ref{eq:sig5.3}) is a consequence of our assumption that there is no "small" length scale in the turbulent velocity field, only the length scales of the flow imposed by the geometry. 

From equation (\ref{eq:sig2}) we have 
 \begin{equation} 
 \sigma_{xz}(z) /\rho = \gamma \vert  \frac{\mathrm{d} u(z)}{\mathrm{d} z} \vert^{1-\alpha} L(z) 
 \label{eq:sig7.1}
 \end{equation} 
 which vanishes at $\vert z \vert=h/2$, as expected from the 
  definition of 
   the stress tensor component $ \sigma_{xz}(z)= \rho < u'_x u'_z>$, as  written  above.
    However  since the vanishing of $ \sigma_{xz}(z)$ close to the wall is not satisfied by the solution (\ref{eq:nu0}) obtained by neglecting the viscous stress, we must include the viscous stress close to the wall. Later we  prove that we have also to include the viscous stress close to $z=0$.
   
Setting $\tilde{z}=h/2-\vert z \vert$  as a local variable (distance to the wall) much  smaller than $h$,  and dividing the two members of (\ref{eq:sig1b}) by $\frac{\vert g \vert h}{2\rho} $  (the r.h.s  value for $z$ close to  $h/2$), we get the following equation
for $ \frac{\mathrm{d} u (\tilde{z})}{\mathrm{d}\tilde{z}} $ 
 \begin{equation} 
\gamma_{1}\vert  \frac{\mathrm{d} u (\tilde{z})}{\mathrm{d}\tilde{z}} \vert^{1-\alpha}  \tilde{z}  +\gamma_{2}\frac{\mathrm{d} u (\tilde{z})}{\mathrm{d}\tilde{z}}=  1
 \label{eq:sigepsilon}
 \end{equation} 
 where  $\gamma_{1} =2 \gamma (d/h- F(h/2))\rho/(gh)$ and $\gamma_{2}=2\rho\nu/(\vert g\vert h)$ is proportional to the viscosity. Introducing the typical velocity $u*$ defined in (\ref{eq:ustar}),  and the scaled velocity $\hat{u}=u/u_{*}$  which is of order unity, equation (\ref{eq:sigepsilon}) becomes
  \begin{equation} 
\hat{\gamma_{1}}\vert  \frac{\mathrm{d} \hat{u} (\tilde{z})}{\mathrm{d}\tilde{z}} \vert^{1-\alpha}  \tilde{z}  +\frac{h}{Re}\frac{\mathrm{d} \hat{u} (\tilde{z})}{\mathrm{d}\tilde{z}}=  1
 \label{eq:sigepsilon2}
 \end{equation} 
where $\hat{\gamma}_{1}$ is also a dimensionless quantity of order unity, defined as $\hat{\gamma}_{1} =2\gamma(\hat{d}/h-\hat{F}(1/2)$, with $\hat{F}$ and $\hat{d}$ given in (\ref{eq:d}) and (\ref{eq:Gz})  respectively, but with $\hat{u}$ in place of $u$.
Let us now consider the junction domain  between the wall and the bulk, where $\tilde{z}$ is much smaller than $h$. In this domain the three terms of (\ref{eq:sigepsilon2}) must have the same order of magnitude.  One has $\mathrm{d} \hat{u} / \mathrm{d}\tilde{z} \sim 1/\tilde{z} $, that gives $\tilde{z}/h \sim 1/Re$. It follows that the first term is  of order unity  if  $ \vert \tilde{z} \vert ^{\alpha} \sim 1$ only if  
   \begin{equation} 
 \alpha = 0
 \label{eq:alf=0}
 \end{equation} 
in the limit of large Reynolds number.

 In summary we find that the exponent $\alpha$ which was free initially, is  actually determined by the  geometry of the  Poiseuille set up, here the  plane boundaries. For $\alpha=0$,
the Reynolds stress  is given by the expression
\begin{equation} 
 \sigma_{xz}(z)/\rho = \gamma \vert  \frac{\mathrm{d} u(z)}{\mathrm{d} z} \vert L(z)
 \label{eq:stressbb}
\end{equation}  
where $L(z)$ is given by (\ref{eq:sig5.2}). In (\ref{eq:Gz}) one can  replace $u(z)-u(0)$ by $u(z)$  because of the translation invariance, reflected here by the fact that only the derivative of the velocity appears in (\ref{eq:sig1b}). Therefore we set
\begin{equation} 
F(z)=u(z)  
  \label{eq:Fb}
\end{equation}    
which implies $F(h/2)=0$. Moreover  $L(z)$ is a linear function of $u$,
\begin{equation} 
L(z)=  \frac{1}{2}  \int_{-z}^{z} {\mathrm{d}} z' u(z') -\frac{z}{h} d
  \label{eq:L2}
  \end{equation} 
with $d$ simply given by the  velocity profile integrated over $z$,
\begin{equation} 
d  =   \int_{-h/2}^{h/2} {\mathrm{d}} z' u(z').
  \label{eq:db}
  \end{equation} 
  
Finally  equation (\ref{eq:sig1b}) becomes
\begin{equation} 
 \gamma \vert  \frac{\mathrm{d} u(z)}{\mathrm{d} z} \vert \, L(z) - \nu \frac{\mathrm{d}u(z)}{\mathrm{d} z} = \frac{\vert g \vert}{\rho} z.
 \label{eq:diffu}
\end{equation}  

Close to $z=h/2$, one has $ L(z) _{z  \to  \pm \frac{h}{2}}  \approx \pm ( \frac{1}{2}-\frac{\vert z \vert}{h}) d$. Using the variable $\tilde{z}=h/2-z$,  equation (\ref{eq:diffu}) becomes
\begin{equation} 
  \frac{\mathrm{d} u(\tilde{z})}{\mathrm{d}\tilde{ z}} ( \gamma \frac{d}{h}\tilde{z} + \nu ) = \frac{\vert g \vert h}{2\rho},
 \label{eq:dubord}
\end{equation}  
which is finally the equation we have to solve close to the wall.

Taking into account that $ u = 0$ for $\tilde{ z} = 0$ (on the wall),  the solution  of (\ref{eq:dubord}) is
  \begin{equation} 
u(\tilde{z}) = \frac{\vert g \vert h^{2}}{2\rho\gamma d}  \ln\left( \frac{ \tilde{ z}  +\frac{\nu h}{\gamma d}}{\frac{\nu h}{\gamma d}}  \right)
 \label{eq:uinner}
 \end{equation} 
where $d $ has the dimension of a  kinematic viscosity, $\gamma $ is dimensionless and 
 \begin{equation} 
 \ell_{\nu}= \frac{\nu h}{\gamma d}
 \label{eq:tildenu}
 \end{equation} 
 is the  width of the boundary layer, which is small  at large Reynolds, see (\ref{eq:Re2}). 
 Note that the solution (\ref{eq:uinner}) has some resemblance with the one derived  within the standard von Karman-Prandtl boundary layer theory, as presented by Landau for instance in section(42) of \cite{LL}. There are however some significant differences. First, (\ref{eq:uinner})  is derived from the analysis of a fully explicit set of equations for the fluid velocity, without any assumption about the value of the Prandtl length. Furthermore this solution is the correct one for our equations all the way from the wall to the region of merging with the outer solution far from the viscous sublayer (see subsection \ref{Matching1} below). This avoids in particular to have, as in Prandtl-von Karman theory, a velocity proportional to the logarithm of  $\tilde{z}$, the distance to the wall, in obvious disagreement with the boundary condition for a viscous fluid. Of course the latter $\ln(\tilde{z})$ dependance could be understood, 
  as done by Landau,  as the local solution valid only in an intermediate range of  $z$ where viscosity is negligible.  But this leaves open the way this solution is matched with the viscous sublayer, something done automatically with our expression of $u (\tilde{z})$ given in  (\ref{eq:uinner}), as shown in the next subsection.

\subsection{Inner-Outer Matching near the wall}
\label{Matching1}

From the boundary layer solution given in equation (\ref{eq:uinner}) one can already guess that the matching of this "inner solution" (i.e. solution near the wall) with the outer solution (solution far from the wall)
 requires some care.  The goal is to obtain same solutions in the matching domain. The outer solution in the domain $\tilde{z}<<h$  is solution of equation (\ref{eq:dubord}) without the viscous stress, namely with $\nu = 0$, which is  of the form
 \begin{equation} 
 u(\tilde{ z}) = \frac{\vert g \vert h^{2}}{2\rho\gamma d}( \ln  \tilde{ z}/\ell  +C )
 \label{eq:outer}
 \end{equation}
where $C$ is a constant, and $\ell \leq h$.
Therefore the solution of this "outer" problem  diverges logarithmically near the wall, and  all the same must match the "inner solution". Both solutions have 
a logarithmic behavior, but  the arguments of the two logarithms differ by a factor of order $1/Re $, that  represents two widely different length scales,  $h$ for the outer solution and  $h/Re$ for the inner solution.
The ratio of the two length scales is however constant. Therefore the corresponding difference between the two expressions for the velocity can be compensated by adding a constant to the outer solution, that is possible because the equation to be satisfied by this outer solution is formally invariant under the addition of an uniform velocity since (\ref{eq:sig2}) involves derivatives of $u(z)$ only. This is, after all, only a reflection of the Galilean invariance (in the $x$ direction)  of  the fluid equations  for an inviscid fluid and perfectly smooth walls.

Said otherwise, the matching between (\ref{eq:uinner}) and (\ref{eq:outer}) is  obtained help to the constant $C$.  
Writing the log-term in (\ref{eq:uinner})  as $\left(\ln((\tilde{z}+\ell_{\nu})/\ell) - \ln(\ell_{\nu}/\ell) \right) $, where $\ell$ is any length, makes clear that  the right choice for the log-term in (\ref{eq:outer}) is to write the parenthesis of  (\ref{eq:outer}) as 
 \begin{equation} 
 \ln  (\tilde{ z}) +\,C\, =\left( \ln  (\tilde{ z}/\ell) - \ln(\ell_{\nu}/\ell) \right).
 \label{eq:C1}
 \end{equation}
 This leaves open the choice of $\ell$. Because this is related to the outer solution, this length scale has to be proportional to the length scale involved in this outer solution, namely h. The following choice 
 \begin{equation} 
C =-  \ln(\ell_{\nu}/h)
 \label{eq:C}
 \end{equation}
  allows to match the outer solution with the inner solution, by  including the large positive constant  $C$  in the two expressions. In summary the solution  (\ref{eq:uinner}) is valid in the whole inner domain close to the wall and at the edge of the outer domain where it matches the solution at finite distance from the wall. 
 Outside of the boundary layer, in principle, the added uniform velocity field (proportional to $C$) makes the dominant part of this velocity field, but this implies that the logarithm of the Reynolds number is large, that can be difficult to achieve in real situations. Nevertheless we notice that the friction coefficient is measured in pipes with very smooth walls. It also decays slowly as a function of the logarithm of the Reynolds number,  as shown by Moody's diagram \cite{Moody}. We discuss below the implications of those remarks for the general problem of turbulent drag on objects of an arbitrary shape.

 \subsubsection{numerical solution}
 \label{sec:num}
 
 The goal is to solve the integro-differential equation (\ref{eq:diffu})  with  boundary conditions at the wall,
 \begin{equation} 
(u)_{\tilde{z}=0} =0  \qquad \textrm{and} \qquad 
 ( \frac{\mathrm{d} u}{\mathrm{d}\tilde{ z}} )_{\tilde{z}=0}  = - \frac{\vert g \vert h}{2\rho \nu}= \,- \frac{u_{*}^{2}}{\nu}
  \label{eq:bc}
 \end{equation}
  First we note that  we can get rid of the parameters in the r.h.s.  of (\ref{eq:diffu}) by scaling  $z$, $\nu$ and the velocity $u$. Defining  the dimensionless quantities $\hat{u}= u/u_{*}$, $\hat{\nu}= \nu /( h u_{*})$ and $\hat{z}= z/h$,   
  (\ref{eq:diffu}) becomes 
    \begin{equation} 
 \gamma \vert  \frac{\mathrm{d} \hat{u}}{\mathrm{d} \hat{z}} \vert \, L(\hat{u},\hat{z}) - \hat{\nu} \frac{\mathrm{d}\hat{u}}{\mathrm{d} \hat{z}}  =  2 \hat{z}
 \label{eq:duscaled}
 \end{equation}
where  
   \begin{equation} 
L(\hat{u},\hat{z})=\int_{-\hat{z}}^{\hat{z}} \hat{u}(z') dz' -\hat{z} d_{\hat{u}}
 \label{eq:Lhat}
 \end{equation}
 with     $d_{\hat{u}}= \int_{-1/2}^{+1/2} \hat{u}(z') dz'$, then  
 $L(\hat{u},\hat{z})$  is still  a linear function of $\hat{u}$.
The latter property allows to notice that (\ref{eq:diffu}) and (\ref{eq:duscaled})
 present a  dilation invariance, which can be stated as follows:  if $u(z)$  is solution of (\ref{eq:diffu})   for a given set of parameters ($\gamma, \nu, u_{*}$), then  $v(z)=\lambda u(z)$   is  also solution of (\ref{eq:diffu})  but with the parameters  $\gamma \to \gamma / \lambda^{2}$ and $\nu \to \nu / \lambda$, although $u_{*}$ keeps unchanged, whatever $\lambda$.  Similarly,  if $\hat{u}(\hat{z})$  is solution of (\ref{eq:duscaled})   for the given set of parameters ($\gamma, \hat{\nu}$), then any  dimensionless function 
   \begin{equation} 
  \hat{ v}(\hat{z})=\lambda\hat{ u}(\hat{z})
   \label{eq:v}
 \end{equation}
 is  also solution of (\ref{eq:duscaled})  but with the parameters  $\gamma \to \gamma / \lambda^{2}$ and $\hat{\nu }\to \hat{\nu} / \lambda$.  The latter property has been found advisable for the iterative numerical  method we have used.
Setting
   \begin{equation} 
 \lambda = \sqrt{\gamma d_{  \hat{ v}}} 
 \label{eq:lambda}
 \end{equation}
 with $d_{  \hat{ v}}=\int_{-1/2}^{+1/2}   \hat{ v}(z') dz'$,   one get the relation between the scaled quantities  $\hat{v}(\hat{z})$ and $\hat{u}(\hat{z})$ and their averaged value (over $z$),
    \begin{equation} 
   \hat{ v}(\hat{z})= \gamma d_{\hat{u}} \, \hat{u}(\hat{z})  \qquad  d_{ \hat{ v}}=  \gamma d_{\hat{u}}^{2} 
 \label{eq:dvdu}
 \end{equation}
 and between the corresponding velocities  $u=u_{*}\hat{u}$, and    $v  =u_{*}\hat{v}$  
   \begin{equation} 
  v(\hat{z})=  \gamma d_{u} u(\hat{z})   \qquad d_{v}  = \gamma d_{u}^{2}
 \label{eq:vu}
 \end{equation}

    \begin{figure}
\centerline{ 
\includegraphics[height=1.5in]{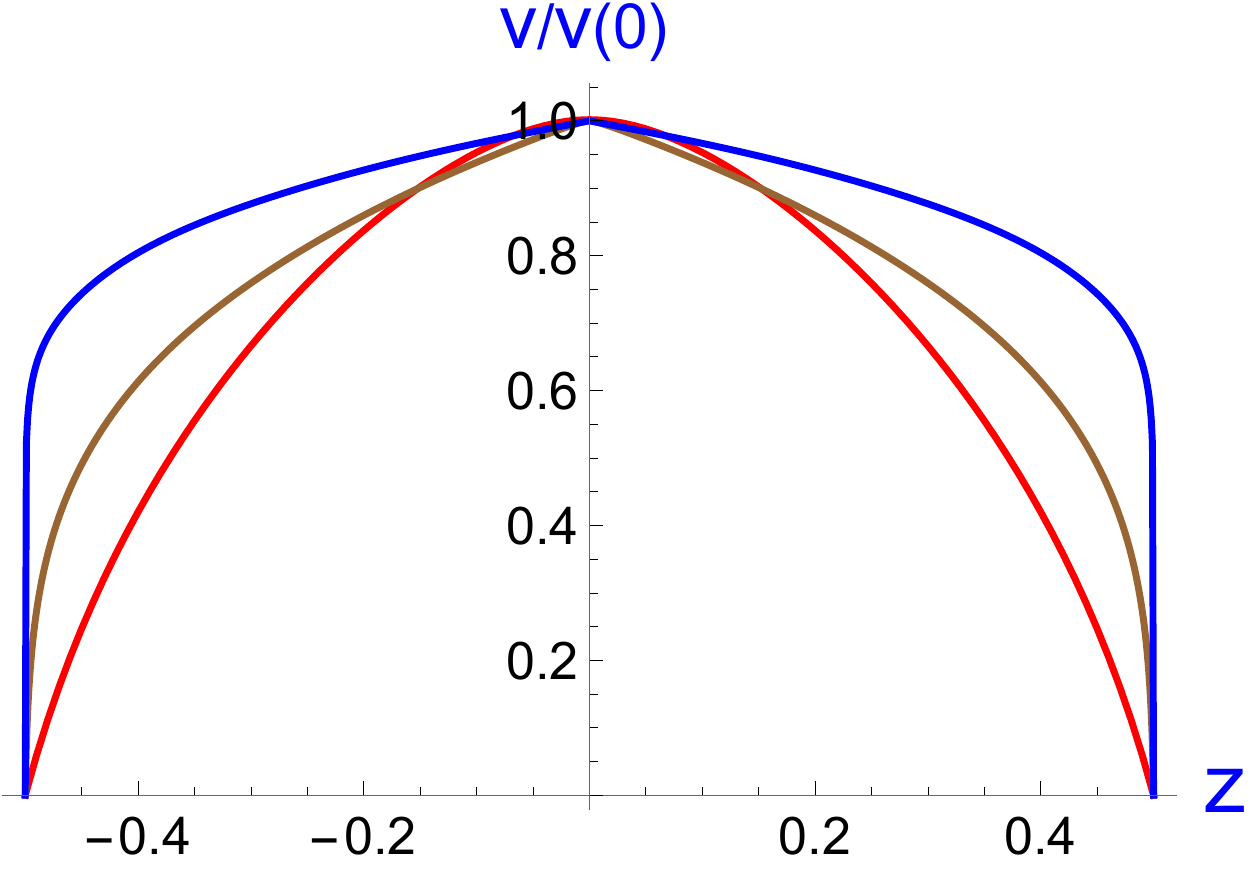}
  }
\caption{ Profile of the scaled velocity $\hat{v}(\hat{z})/\hat{v}(0) = u(\hat{z})/u(0)$, 
 versus $\hat{z}=z/h$,    along the direction $z$ perpendicular to the two plates, for $\nu_{\hat{v}}= 10^{-1}$,  (red)\; $10^{-3}$ (brown) and $10^{-8}$ (blue) 
 or if $\gamma=1$, $Re=10, \; 10^{3}$ and $10^{8}$, see (\ref{eq:Re2}).  At small Reynolds number the profil is round at the top, although a wedge-like top is formed at the top for large Reynolds numbers, yet present on the brown curve (not clear on the figure).}
\label{fig:profile}
\end{figure}

    \begin{figure}
\centerline{ 
(a)\includegraphics[height=1.5in]{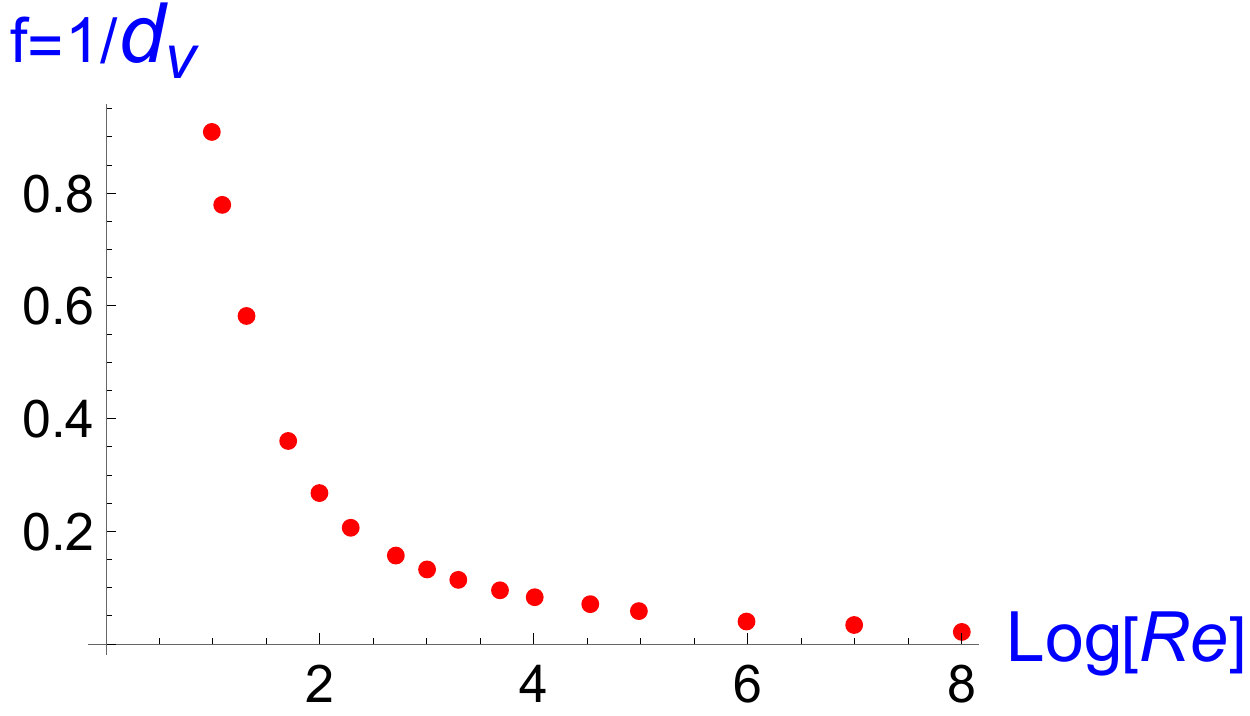}
}
\centerline{ 
(b) \includegraphics[height=1.5in]{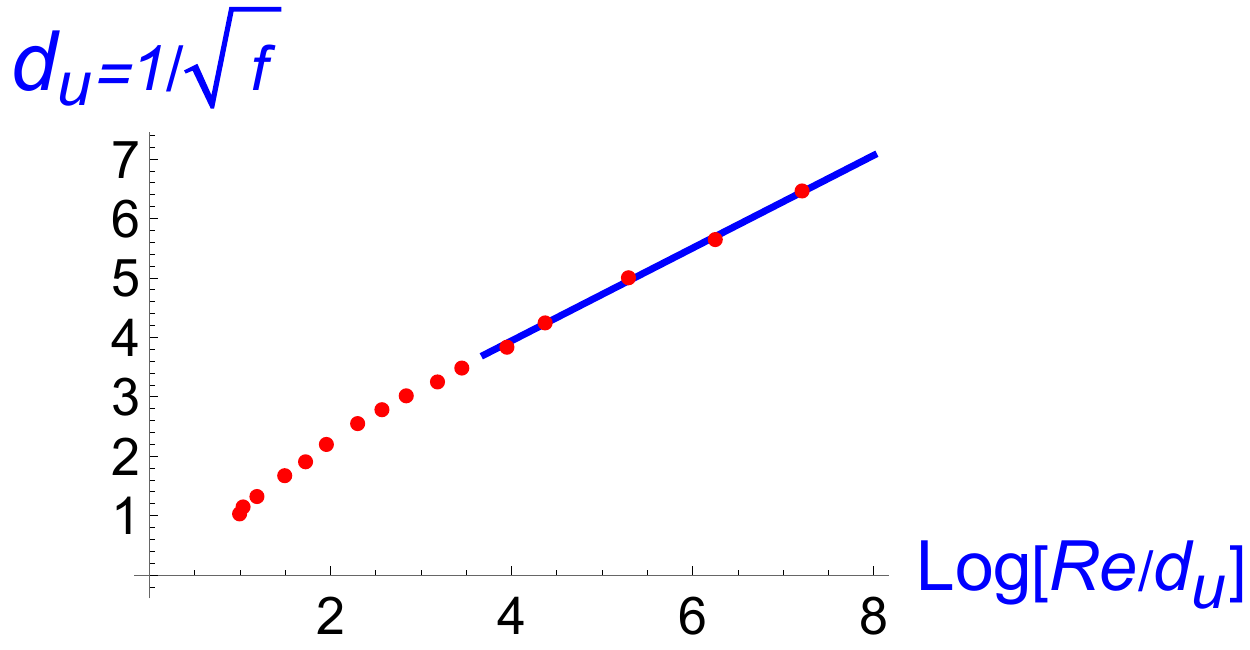} 
(c) \includegraphics[height=1.5in]{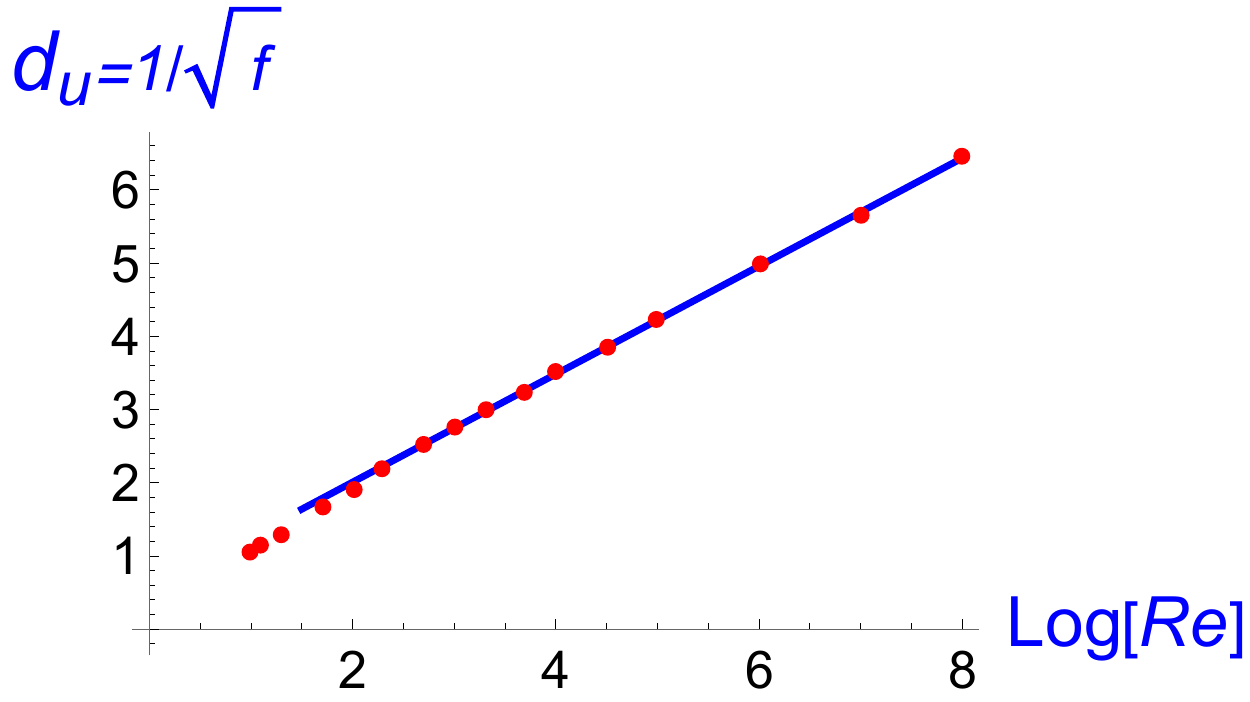}
  }
\caption{Friction coefficient defined here as $f=1/d_{\hat{v}}$,  ($d_{\hat{v}}$ given in  (\ref{eq:dvdu}) is the mean value  of $\hat{v}$ (averaged over the whole profile), and is equal to $d_{\hat{u}}^{2}$).  (a) : $1/d_{\hat{v}}$  is plotted versus $\log_{10} (\nu_{\hat{v} }) $  (or $\log_{10} (Re)$\, if $\gamma=1$). It reflects the behavior  of  $\Lambda_{x}$, defined in (\ref{eq:frict}), as $Re$ increases. 
(b) and (c) :  $1/\sqrt{f}$ versus $Log_{10}(Re \sqrt{f})$  and versus $Log_{10}(Re)$ respectively.   The blue lines are linear fits. Note that  both curves (b) and (c), seem to increase linearly,  although the abscissa slightly differs, this is because the difference is about $\sim \log(\log Re)$). The von-Karman law  in (b) is expected to be valid for $ Re$ larger than $10^{5}$.  
}
\label{fig:logRe}
\end{figure}

   \begin{figure}
\centerline{ 
\includegraphics[height=1.5in]{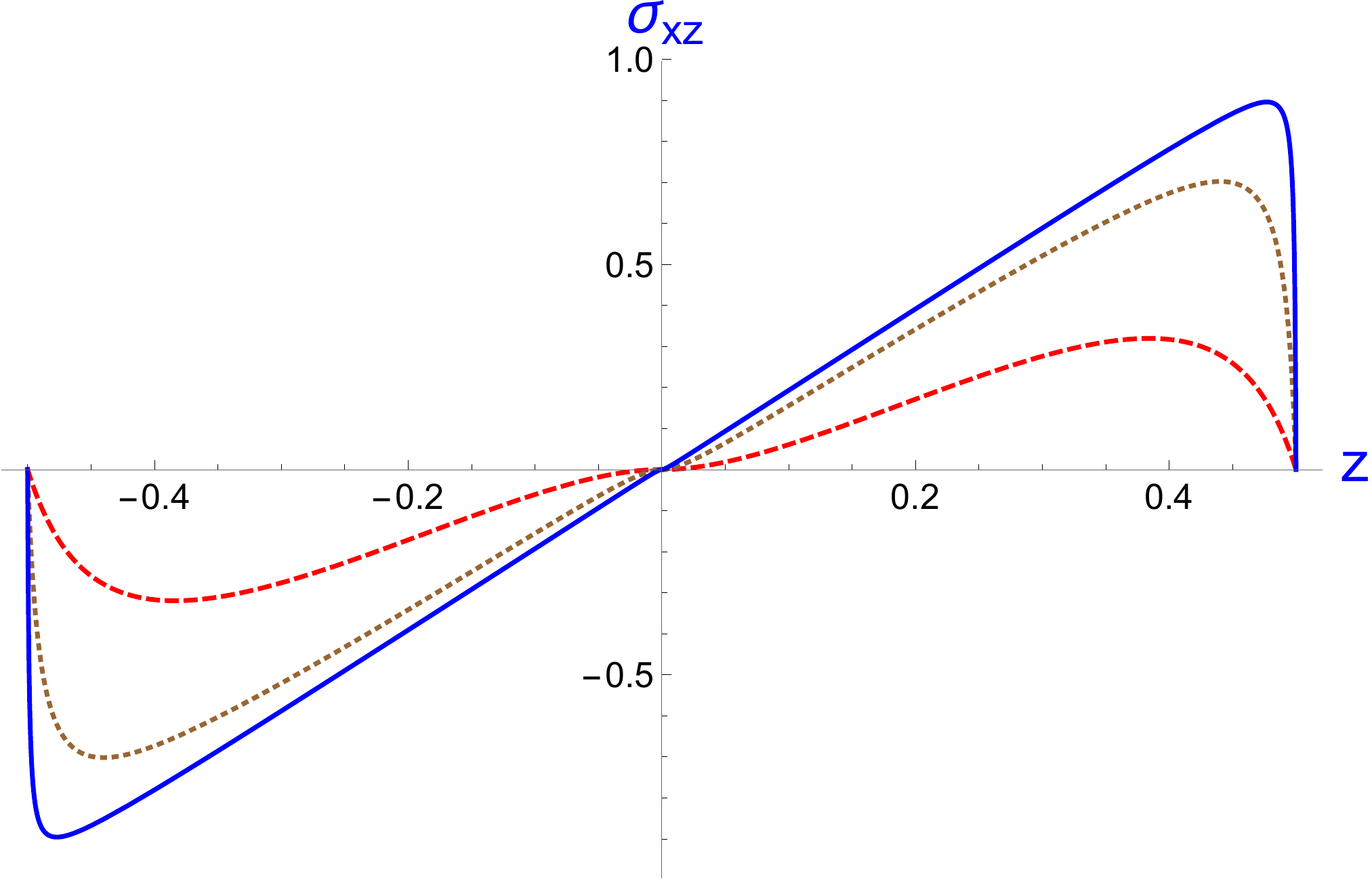}
  }
\caption{ Profile of the Reynolds stress component $\sigma_{xz}$,
 versus $\hat{z}=z/h$,  along the direction $z$ perpendicular to the two plates, for $\nu_{\hat{v}}= 10^{-1}$ (red, dashed),\; $10^{-2}$ (brown, dotted) and $10^{-3}$ (blue, solid), 
 or if $\gamma=1$, $Re=10, \; 10^{2}$ and $10^{3}$.  As $Re$ increases, the profile becomes linear in the bulk, except on the edges (boundary layers) where it falls down to zero, more and more abruptly.
 }
\label{fig:sigma}
\end{figure}

  The integro-differential  equation (\ref{eq:duscaled}) becomes in terms of the dimensionless function $\hat{ v}(\hat{z})$, 
    \begin{equation} 
 \frac{\mathrm{d}   \hat{ v}}{\mathrm{d} \hat{z}} 
 \left(  \frac {1}{d_{  \hat{ v}}} \vert L(  \hat{ v},\hat{z} \vert 
 + \nu_{\hat{v}} \right)   = -\, 2  \hat{z}
 \label{eq:newdu}
 \end{equation}, 
 where L(.) is defined in (\ref{eq:Lhat}) and $ \nu_{ \hat{v}}= \hat{\nu}/ \sqrt{ \gamma d_{\hat{v}}}$. Putting (\ref{eq:vu})  into the definition of the Reynolds number $Re= d_{u}/ \nu $, we recognize the inverse of the  Reynolds number,  replacing the one estimated above in (\ref{eq:Re}), 
 \begin{equation} 
 \nu_{  \hat{ v}}=  \hat{\nu}/ \sqrt{ \gamma d_{\hat{v}}}=  \frac{1}{\gamma Re}
 \label{eq:Re2}
 \end{equation}
   
Help to the change of variables $u \to  v= \lambda u$, the iteration method  has shown up very efficient. We
start from an even log-profile which fulfills the boundary conditions  ( \ref{eq:bc}) at the wall. For the scaled velocity $\hat{v}$, these conditions are
 \begin{equation} 
\hat{v}(\hat{z}=1/2)=0     \qquad \textrm{and} \qquad 
 ( \frac{\mathrm{d}   \hat{ v}}{\mathrm{d}\hat{ z}} )_{\hat{z}=1/2}= - \frac{1}{\nu_{  \hat{ v}}}
  \label{eq:bc}
 \end{equation}
  We have  chosen the simple test function for the initial  profile 
 \begin{equation} 
  \hat{ v}_{test}( \hat{z}) =\beta \ln \left(1+k(\frac{1}{4}-\hat{z}^{2}) \right)
 \label{eq:test}
 \end{equation}
which vanishes at the wall. Close to the boundary $\hat{z}=1/2$  the parenthesis in (\ref{eq:test}) can be approximated by its first order expansion  in terms of  the scaled variable $\tilde{z}=1/2-\hat{z}$  (as above but in scaled form),
 \begin{equation} 
  \hat{ v}_{test}(\tilde{ z}) \approx \beta \ln (1+k \tilde{z})
 \label{eq:test1}
 \end{equation}
 and the condition (\ref{eq:bc}) for the derivative yields
  \begin{equation} 
\beta k=\frac{1}{\nu_{  \hat{ v}}}
 \label{eq:bc2}
 \end{equation}

A second relation between the two parameters $\beta$ and $k$  can be deduced by  looking at  the mid-plane velocity. From   (\ref{eq:test}) we have $\hat{v}_{test}(0)=\beta \ln(1+k/4)$.  The outer solution  for $  \hat{ v}(\hat{z})$, is  the  solution of   (\ref{eq:newdu}) without the viscous term and for $\tilde{z}$ small, as  performed above for the outer solution of $u(z)$, see (\ref{eq:outer}) and (\ref{eq:C}). Taking into account the successive rescaling, we  get the outer solution $\hat{v}(\tilde{z})= \left( \ln  (\tilde{ z}/\ell) - \ln(\ell_{\nu}/\ell) \right)$. Extending this solution until the center of the flow, one may fit the outer solution with the test function by  setting  $\ell=h/4$, and 
$  \hat{ v}(\tilde{z}=1/2)  \, =1 $, or $\beta=1$. Finally, using (\ref{eq:bc2}), the profile of the scaled velocity $  \hat{ v}(\hat{z})$
 is  obtained by iterating  the solutions of (\ref{eq:newdu})  and using the test function 
 \begin{equation} 
  \hat{ v}_{test}( \hat{z}) =   \ln \left(1+(\frac{1}{4}-\hat{z}^{2})/\nu_{  \hat{ v}}\right).
 \label{eq:test2}
 \end{equation}
The convergence of the solution occurs very rapidly for  $\nu_{\hat{ v}}$ values of order $10^{-1}-10^{-2}$, but needs more  precision machine and more steps as $\nu$ decreases. This is due to the behavior of the solution which becomes more and more stiff near the walls.

The profile of  the velocity $u(\hat{z})/u(0)$,  or $\hat{v}(\hat{z}) /  \hat{v}(0)$,  is shown in Fig.\ref{fig:profile}  for  $\nu_{  \hat{ v}}$ decreasing from $10^{-2}$ to $10^{-8}$. For $\nu_{  \hat{ v}}$ smaller than $10^{2}$ the profile displays 
 a wedge-like form in the bulk  associated with  a  strong gradient near the walls. 
 As  $Re$ increases (or $\nu_{  \hat{ v}}$ decreases) the bulk profile enlarges and the slope at the wall increases in agreement  with (\ref{eq:bc}).  
 
 In the same range of Reynolds number, we plot in Fig. \ref{fig:logRe}-(a)  the inverse off   $1/d_{\hat{v}}$  defined in equation (\ref{eq:dvdu}), which  
  is proportional to the friction coefficient,  because  $d_{\hat{v}} = \gamma d_{\hat{u}}^{2}$.  The curve $1/d_{\hat{v}}$  decreases with respect to $\ln(Re)$, as expected.   Figs.  \ref{fig:logRe}-(b)-(c) 
   complete the study of the friction behavior, see the discussion in Sec.\ref{turbulentdrag}.   
 
The profile of the stress tensor $\sigma_{xz}$ is drawn in Fig.\ref{fig:sigma}   for $\nu_{  \hat{ v}}$ decreasing from $10^{-1}$ to $10^{-3}$.  As $Re$ increases, the Reynolds stress tends to a linear function of $z$ in the bulk, which falls down  more and more abruptly at the boundaries.

We have two remarks, the first one concerns  the Reynolds number.  In (\ref{eq:Re}), $Re$  was estimated  by using $u*$  defined in (\ref{eq:ustar}) as the order of magnitude of the velocity in the bulk, namely without the effect of the boundary layer.  However this value $u_{*}$ is noticeably  underestimated when viscosity effects are included, especially in the limit of $\nu$ tending to zero.   This is due to the effect of the added constant $C$, see (\ref{eq:C}).  The actual mean velocity in the flow is much better represented by   $ d_{u}= u_{*}h d_{\hat{u}}$ which increases  approximately as  $(\ln (1/ \nu_{\hat{v}}))^{1/2}$ at large $Re$, that justifies the definition of  the Reynolds number as $Re= d_{u} /\nu$  in (\ref{eq:Re2}). 

A second remark  concerns the link between $d_{\hat{u}}$ and $d_{\hat{v}}$. These quantities  are the mean value  of $\hat{u}$  and $\hat{v}$ respectively (averaged over the whole profile).
   We showed above that $d_{\hat{v}}$  is equal to the square of  $d_{\hat{u}}$ times $\gamma$, see (\ref{eq:dvdu}).  Therefore one finds that the decrease of  the friction factor (proportional to $1/d_{u}^{2}$ ) with respect to $\ln(Re)$, agrees qualitatively   with the observations in the case of smooth walls\cite{Moody}.
   
    This difference of behavior with respect to the physical parameters reflects, in our problem, the very strong heterogeneity of the velocity field.

 \subsection{Inner-Outer Matching near the middle of the flow}
\label{Matching 2}

   \begin{figure}
\centerline{ 
(a)\includegraphics[height=2in]{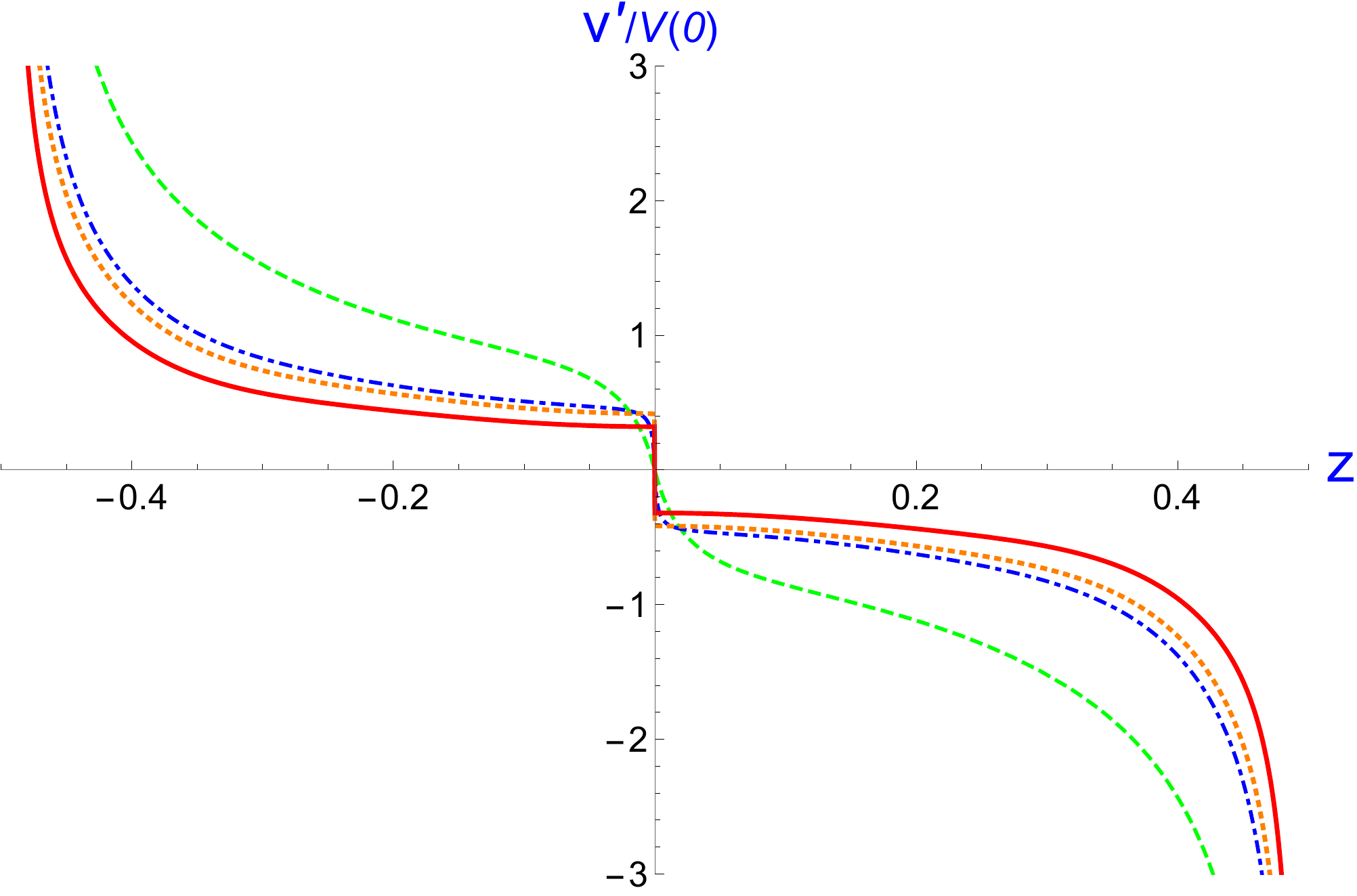}
(b) \includegraphics[height=1.75in]{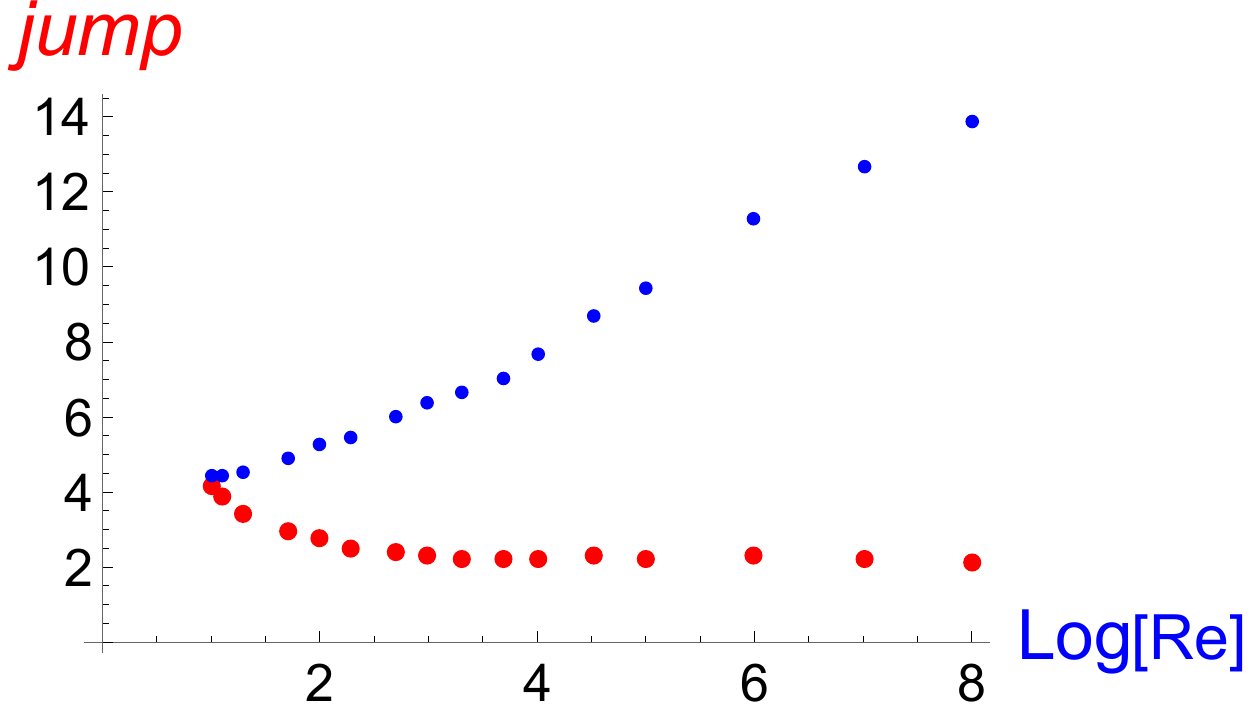}
 }
\caption{
 (a) Profile of the  velocity  gradient  in the central plane (close to $z=0$), scaled to the central velocity. 
  $\frac{ \mathrm{d} \hat{ v}}{\mathrm{d} \hat{ z}}   / \hat{ v}(0)$
 versus $\hat{z}$, for $\nu_{\hat{v}}= 10^{-2}$ 
(green  dashed line), \, $10^{-4}$ (blue, dotted-dashed line),\, $10^{-6}$  (orange dotted line) \, and $10^{-8}$ (red solid line),
 or  if $\gamma=1$, $Re=10^{2}$,  $10^{4}$, $10^{6}$ and $10^{8}$ respectively.   Close to $z=0$, the slope  of the derivative increases with the Reynolds number, 
 then  evolves smoothly in the bulk.  
 (b) Jump of the velocity gradient, $\lbrack \frac{ \mathrm{d} \hat{ v}}{\mathrm{d} \hat{ z}}(0_{+})  -\frac{ \mathrm{d} \hat{ v}}{\mathrm{d} \hat{ z}}(0_{-}) \rbrack / \hat{ v}(0)$. The upper curve (blue points) displays the jump of $\frac{ \mathrm{d} \hat{ v}}{\mathrm{d} \hat{ z}}$, which  increases with $Re$, the lower curve shows the jump of $\frac{ \mathrm{d} \hat{ u}}{\mathrm{d} \hat{ z}}$,  equation (\ref{eq:saut}),  which decreases first and tends to an asymptotic constant value.
 }
\label{fig:profderv}
\end{figure}

As we discovered when looking for a numerical solution of the equation of momentum balance, there is another boundary layer near the center of the flow, that is near $z = 0$. We do not believe this is an artefact of our modeling of the turbulent stress. The occurrence of this boundary layer can be explained as follows. Because we assumed that the average velocity $u(z)$ is an even function of $z$, it reaches a maximum at $z= 0$. Supposing this is a smooth maximum, it is also a place where the fluid velocity is almost uniform. Therefore this is a place where, physically, the source of turbulence is absent: no turbulence is created by a uniform flow. Therefore the balance of momentum can become dominated by viscous friction. This explanation is obviously quite approximate, but is is supported by a detailed analysis of the solution of equation (\ref{eq:sig1b}) near $z = 0$ in the limit  $ \nu \to 0$. The analysis of the solution in this limit uses the same equation as in section \ref{Matching1}, but with different boundary conditions. 

 The Reynolds stress $ \sigma_{xz}$, given by equation (\ref{eq:stressbb}),
 is an odd function of $z$,   $\vert  \frac{\mathrm{d} u(z)}{\mathrm{d} z} \vert$ is an even function of $z$,  and $L(z)$ is an odd function of $z$. Near $z = 0$ one can approximate $L(z)$ by the first non vanishing term in its Taylor expansion, $L(z)\approx\beta z$, with $$\beta_{u}= u(0)-d_{u}.$$
 Notice that the parameter $\beta_{u}$ depends on the full velocity field, not on  the local velocity only. Plugging the latter approximation of $L(z)$ in equation  (\ref{eq:stressbb}) one finds the following ordinary differential equation for $u(z)$ near $  \vert  z  \vert = 0$
\begin{equation} 
 \beta_{u} \gamma z \vert  \frac{\mathrm{d} u(z)}{\mathrm{d} z} \vert -  \nu \frac{\mathrm{d} u(z)}{\mathrm{d} z} = \frac{\vert g \vert  z}{\rho} 
 \label{eq:sig7.2}
 \end{equation} 
This equation can be integrated at once, with the result 
  \begin{equation} 
u(0) -u(z)=  
\frac{\vert g\vert }{\rho (\beta_{u}\gamma)^{2}} \left(  \beta  \gamma \vert z \vert   - 
\nu  \ln (\frac{\nu + \beta_{u}  \gamma \vert z \vert  }{\nu}) \right)
 \label{eq:sig7.4m}
 \end{equation}

This velocity $u(z)$ changes on a distance of order $\delta z =\nu/(\beta_{u} \gamma)$ from a positive to a negative derivative. On this distance  the jump of  the derivative, labelled $\delta u_{,z}$, is 
 \begin{equation} 
 \delta u_{,z} = \vert g \vert/(\rho\gamma \beta_{u})
\label{eq:saut}
 \end{equation} 
  a quantity which  could depend  of  the viscosity via $\beta_{u}$. Nevertheless Fig.\ref{fig:profderv} shows that the jump  of $du/dz$ tends to a constant in the limit of large $Re$. Differently the jump of $dv/dz$, $\delta v_{,z}= \lambda \delta u_{,z}$ increases  with $Re$ (upper curve with blue points).
 It is interesting to notice that the jump of slope of  the velocity $u(z)$ 
 is not a tangential discontinuity, it is like a localized jet in the middle of the layer.

\subsection{$\sigma_{ii}$ components of the Reynolds stress tensor }

The above study of  the turbulent plane  Poiseuille flow was restricted to the  role played by the component $\sigma_{xz}$ of the Reynolds stress.
From the relationship ( \ref{eq:sig0}) between turbulent stress and velocity fluctuations we infer that, besides  $\sigma_{xz}$, other components of this stress  do not cancel and so should be considered for possible effects on the average velocity. For instance  the diagonal components, $\sigma_{xx}=\rho <u'^2_{x}>$, $\sigma_{yy}=\rho <u'^2_{y}>$  and $\sigma_{zz}=\rho <u'^{2}_{z} >$ are generally non zero because they are averages of squares.
The model   of the turbulent Reynolds stress described in \cite{2}  includes, in addition to the term ( \ref{eq:sig11}) studied above, the following  diagonal tensor,
 \begin{equation} 
\sigma_{ij}({\bold{x}}) =  \delta_{ij} \gamma_{i} \rho  \vert {\bold{\nabla}}\times {\bold{u}}({\bold{x}}) \vert ^{1-\alpha }\int {\mathrm{d}} {\bold{x}}'  \,{\bold{K}}({\bold{x}},{\bold{x'}}) \,  \vert {\bold{\nabla}}\times {\bold{u}}({\bold{x'}}) \vert  ^{\alpha + 1} 
 \label{eq:sigdiag}
\end{equation}
where $\delta_{ij}$ is the Kronecker symbol, and the constants $ \gamma_{i}$  are introduced to fulfill the Schwartz inequalities  for the correlation functions $<u'_{i}u'_{j}>$, named realizability conditions, see \cite{sch}.
For the case $\alpha=0$ treated here, the latter expression becomes
$\sigma_{ii}({\bold{x}}) =    \gamma_{i} \rho  \vert {\bold{\nabla}}\times {\bold{u}}({\bold{x}}) \vert \int {\mathrm{d}} {\bold{x}}'  \,{\bold{K}}({\bold{x}},{\bold{x'}}) \,  \vert {\bold{\nabla}}\times {\bold{u}}({\bold{x'}}) \vert$, 
which gives for our set up, 
\begin{equation} 
 \sigma_{zz}(z)/\rho = \gamma_{z} \vert  \frac{\mathrm{d} u(z)}{\mathrm{d} z} \vert L(z),
  \label{eq:sigdiagb}
\end{equation}
Let us notice that in the end, the diagonal component of the RST is  proportional to  the non diagonal term (\ref{eq:sig7.1}),   meanwhile the non diagonal  component is derived from (\ref{eq:sig11}) although  the  expression (\ref{eq:sigdiagb}) was  derived from (\ref{eq:sigdiag}). We can show that the diagonal components in (\ref{eq:sigdiagb}) do not contribute to forces driving the Poiseuille flow.
In such case the   diagonal tensor $\sigma_{ii}$  can be seen as  a time averaged pressure depending on the spatial coordinates, and both quantities, pressure and  diagonal tensor,  are impulse carrier. 
As written in \cite{2},  the couple $({\bold{u}},p)$ is not unique in dynamical incompressible systems, since $p$ is  a scalar  jauge field defined up to an additive scalar function. In other words the  pressure is not  an independent variable, but a Lagrangian multiplier necessary to ensure the incompressibility, since it fulfills the relation $ \Delta p =u_{i,j}u_{j,i}$ with summation over the same indices.  

In order to prove that the diagonal components in (\ref{eq:sigdiagb}) do not contribute to forces driving the Poiseuille flow,  let consider for  instance the balance of forces in the $z$ direction,
 \begin{equation} 
 \frac{\mathrm{d}}{\mathrm{d} z}(\rho  u_{z}^2 +  \sigma_{zz} + p(z)) = 0 
  \label{eq:uz}
\end{equation}  
where $ u_{z}$ is a hypothetical component of the average velocity in the $z$ direction, and $p(z)$ is a  scalar function which could  also depend on  $z$.

One may set  $p(z)=\,- \, \sigma_{zz}(z)$,  
that cancels the contribution of  the diagonal term $\sigma_{zz}$ to the forces along $z$, and leads to $u_{z} =0$,
so that the balance of forces in the $z$ direction is realized without flow velocity in this direction. 
Note however that all this relies on the validity of the assumption that  all quantities (like the  Reynolds stress tensor) depend on $z$ only, which is relevant
in the case of pipe flows of uniform cross section  along $x$ (those with a translation invariance in the direction perpendicular to this cross section). It does not imply however that solutions depending on $z$-only are the only possible ones. Consider for instance a cross section made of two circles connected on a finite segment near the axis of their centers, and suppose the two circles of different diameters. There shall be less friction on the wall for the largest circle and so, for the same pressure head, the speed  is bigger in the biggest circle. This will yield a shear layer at the juncture of the two circles and therefore an instability generating a dependance of the average velocity with respect to the variables $y$ and $x$. Such an undulation will likely give a fairly complex average velocity field.

\section{ Skin drag versus turbulent drag}

\label{turbulentdrag}

The turbulent Poiseuille flow makes a standard exemple of turbulent flow  with skin drag due to boundary layers along the solid surface,  and shear-generated turbulence away of it. Such shear turbulence was also considered in reference  \cite{2} in the mixing layer behind a splitter plate, a situation quite different of the Poiseuille flow, because the splitter plate undergoes a lift force due to the turbulent drag behind it. Therefore it is of interest to look for  general conclusions to be drawn in situations where lift forces exists or not.
Of course we think first to the case already considered by Newton in  the Principia, namely the one of a blunt body moving quickly in a fluid immobile far from it. Newton showed that the drag felt by this object grows like the square of its velocity with respect to the fluid at rest at infinity. Neglecting vector indices, Newton's law of drag is 
\begin{equation} 
 F_N  = - C_x \rho S  \vert U \vert U
   \label{eq:sig9}
 \end{equation} 
where $S$ is the cross section of the object moving at speed $U$ and $C_x$ is a numerical constant depending only on the shape of the moving body. The absolute value there is to recall that the drag force $F_N$ changes sign as the velocity is reversed (a non trivial property of turbulence related to dissipation). Newton's formula is remarkable in many ways. Some comments would deserve to be  made.

 The experiments show that the coefficient $C_x$ may have a complex behavior as the Reynolds number increases, depending on the shape of the body. In the  case  of spheres this is well known as the "drag crisis" 
  At increasing speeds the drag coefficient jumps by a finite amount and tend to a  non-zero constant value at  fairly large values of $Re$,  a value that is practically insensitive to the roughness of the surface of the blunt body.  For Poiseuille-typed flows,  and more generally for pipe flows of arbitrary cross section, one defines a friction coefficient $\Lambda_x$ which is
 more or less equivalent  to $C_{x}$. It is
the drag force $D_r$ per unit length of a turbulent pipe flow of velocity $U$ in a pipe of diameter $h$, by the relation
\begin{equation}
 D_r  = - \Lambda_x  \rho  \vert U \vert U h 
   \label{eq:frict}
 \end{equation} 
The Ch\'ezy coefficient $\Lambda_x$ depends on Reynolds number, and displays a sensitivity to the smoothness of the surfaces.  Experimentally $\Lambda_{x}$  decays approximately like $1/ \ln (Re)$ at large $Re$,  for flows in pipes with very smooth surfaces,  as it follows from our study of  turbulent Poiseuille flow, see Fig.\ref{fig:logRe}. 

A significant question posed by real pipe flows for a long time, is the effect of the roughness of the solid surfaces. This is obviously related to the behavior of the friction coefficient at very large values of the Reynolds number, because the thickness of the viscous boundary layer is of order $h/Re$ and so can become 
in real flows 
 of the order of magnitude of the mean height of the rough  surface, $\ell_{ru}$, defining the
 length scale of  the roughness of the wall.  A simple argument shows that  the length $ \ell_{ru}$  should play the same role as our expression for the thickness of the viscous sublayer, $ \ell_{\nu}$. 
 Let us assume, as observed, that the velocity vanishes in a small domain close to the surface, $ 0 < \tilde{z} < \ell_{s}$, where $\ell_{s}= k \ell_{ru}$  is a fraction of the roughness length,  the coefficient $k$ depending on various parameters like the orientation of the asperities with respect to the main flow.  The integration of (\ref{eq:dubord}) between $\ell_{s}$  where the solution est zero, and $\tilde{z}$
 leads to the solution 
  \begin{equation} 
u(\tilde{z}) = \frac{\vert g \vert h^{2}}{2\rho\gamma d}  \ln\left( \frac{ \tilde{ z}  + \ell_{\nu}}{\ell_{s}+\ell_{\nu}} \right)
 \label{eq:ubord2}
 \end{equation} 
with $\ell_{\nu}=\frac{\nu h}{\gamma d}$, as in (\ref{eq:tildenu}). 
It follows that the constant $C$ appearing in the outer solution in (\ref{eq:C})  becomes 
  \begin{equation} 
C =- \ln ( \ell_{s} +\ell_{\nu}).
 \label{eq:C2}
 \end{equation}
  The important difference  between $\ell_{s}$ and $\ell_{\nu}$,  lies in the fact that $\ell_{s}$ does not depend on the Reynolds number.  
 This leads to the conclusion that the velocity of the plug flow to be added to match the boundary layer due to the roughness of the surface of the pipe is of order $( g h/\rho)^{1/2} \ln(\ell_{s}/h)$ in the limit of very large Reynolds number,  so that the friction factor, instead of tending to zero as  $1/\sqrt{f} \sim \ln (Re)$ (in the case of smooth surface),  tends now to a small constant such that
\begin{equation}
(1/\sqrt{f})_{Re \to \infty} \sim \ln (\ell_{s}/h).
    \label{eq:frictrug}
 \end{equation} 
The relation (\ref{eq:frictrug}) agrees with  the Colebrook  relation  \cite{Moody},  $1/\sqrt{f} =-2 \log_{10}(a\,  \ell_{s}/h + b /Re\sqrt{f})$ where $a$ and $b$ are constant factors depending on the geometry of the flow, that  extends the Von Karman-Prandtl expression (same with $a=0$), to the case of rough surfaces.

Because the skin drag is sensitive to the structure of the wall at the scale of the thickness of the viscous boundary layer, it seems possible to change the contribution of the exchange of momentum between the fluid and the wall, by tuning the structure of the roughness of the wall, for instance by undulations of this wall in the range of length scales of order of the viscous sublayer. This would introduce anisotropy in the balance of momentum at this scale and so at bigger scales of the flow, beyond the boundary layer.

Without claim of a rigorous theory for those different behaviors of $C_x$ and $\Lambda_x$ at large Reynolds number, we outline below an explanation based partly on the present study. 
 In  turbulent Poiseuille  flow,   the drag comes only from the viscous friction on the wall  (leaving aside the central jet in the middle of the bulk,  see Sec. \ref{Matching 2} ), related itself to the component $\sigma_{xz}(z)$  of the  Reynolds  stress there. But this turbulent stress is null on the wall, then it does not act directly on it, it acts  only as a boundary value for the viscous layer. As shown above,  the abrupt drop of $\sigma_{xz}(z)$ close to the wall is possible help to the viscosity, which enters into play  
in  the interconnection domain between the inner/outer solution which is of  $log(Re)$ form, see (\ref{eq:C}), that  allows to transfer the $\sigma_{xz}$ component across the boundary layer where it becomes viscous.
 This explains why the logarithm of the Reynolds number appears  
  in the friction factor which decays to zero at large $Re$.
  
  This (the $ \log Re$ term in the solution close to the wall) remains true whenever all non diagonal components  $<u'_{i}u'_{j}>$ vanish at the boundaries, which occurs in channel and pipe flows, and more generally 
whenever $i$ and/or $j$ is a local Cartesian component of the velocity fluctuation in the direction normal to the wall.

On the other hand, in flows with a general structure, like around a blunt body at large speed,
 there is another contribution to the stress, besides turbulent Reynolds  stress, which is the isotropic pressure. Let us recall that D'Alembert  paradox states that an inviscid incompressible flow around an obstacle exerts no drag because of the exact balance between the upstream and downstream contribution to the force  coming from the pressure, that is in direct contradiction to the observation of substantial drag on bodies moving relative to fluid.  
 
 The explanation of this paradox is that first the time dependent Euler equations in 3D do not make in general a well-posed problem, which needs some sort of regularization of small scales, either by viscosity or other effects like radiation of sound waves because of the compressibility.  Once the solution is regularized the singularities are transformed into 
local dissipation events, something coherent with our representation of the Reynolds stress tensor by an integral in space of a quantity quadratic with respect to the average velocity field. Of course the local dissipation by singular events only concerns the  flow  far from the viscous boundary layer.

 In the case of the turbulent drag due to pressure difference between the two sides of the obstacle, a pressure of  order of magnitude  $\rho u^2$, the drag force due to it, takes the form given by Newton's relation ( \ref{eq:sig9}),  without dependance on $Re$ (asymptotically  at large $Re$) because the $Re$ dependance  concerns the skin force, an auxiliary  boundary layer problem with viscosity.  
 Note that  the absolute value in the equation (\ref{eq:sig2})  which breaks the symmetry under velocity reversal, is necessary for the description of both phenomena, the skin drag and the turbulent drag.

One may draw some conclusions from the way the inner-outer matching is done near the wall in the case of  the turbulent Poiseuille flow. As we have just shown, the matching  links the coefficient  $C=-\ln (\ell_{\nu}/h)$ of the logarithmic dependance of the solution,  with 
the amplitude of a uniform mean velocity $d_{u}/h$, solution of the equation in the turbulent domain. The addition of the constant solution $C$ arising from the boundary condition, is possible in our model where the turbulent stress depends on the vorticity,  therefore  is not affected  by the addition of a  constant velocity. 
Another point of interest raised by our approach of the structure of the average velocity field of turbulent flows, is the possibility of bifurcations of the solution in the limit of a very large Reynolds number. This kind of bifurcation is known to occur in such flow,
 for instance in the wake of fast moving cars, where this wake looses its left/right symmetry and so yields an unwanted torque on the car \cite{cadot}. Because our equation for the averaged velocity field is non linear, its solutions may break symmetries like the left/right symmetry of some wakes.
We plan to look at the questions quoted above in future works. 

 \section{Conclusions and perspectives}
  \label{sec:Concl}
  
  In this paper we wrote fully  explicitly the integral equation for the balance of momentum including the closure of the turbulent stress introduced in ref  \cite{1} on the basic assumption that dissipation  is caused by singular events described by solutions of Euler's equation. Because of the fully explicit character of this closure, it is possible to obtain results more detailed than what was derived long ago by Prandtl and Landau.  Particularly we obtain  the log-law  close to the wall, in a so to say rational way, that is by handling explicit equations from the beginning.
   Our analysis displays also a clear matching between the boundary layer solution, valid close to the wall, and the outer solution far from the wall. Roughly speaking our integral representation of the turbulent stress is a way to get rid of an impossible definition of Prandtl mixing length far from the boundary layer (the one close to the wall). Thanks to that we have the full solution, everywhere  between the two walls,
  and related to a unique equation. This allows to understand 
   how the solution far from the wall tends in the limit of a very large Reynolds number to an almost uniform velocity plus an (in principle small) added velocity of amplitude of the order of the inverse of the logarithm of the Reynolds number, a quantity small in principle in the limit of a very large Reynolds number, 
   although likely very hard to achieve in real experiments. This occurrence of logarithms in the skin friction explains 
    that the friction factor in pipes and channels with smooth surfaces, tends to decay  to zero, qualitatively as a function of the logarithm of the Reynolds number,  and also that the drag coefficient of blunt bodies is insensitive  to $Re$  when their surfaces are perfectly smooth, but  are sensitive to $Re$, including at large values of it,  when their surfaces are rough.

\section*{Acknowledgement}
We thank  Christophe Josserand for his interest in this work and for fruitful and stimulating discussions.

  \end{document}